\documentstyle[twoside,fleqn,epsf,espcrc2]{article}

\newcommand{\bda}{\begin{\displaymath}\begin{array}{rl}}
\newcommand{\eda}{\end{array}\end{displaymath}}
\newcommand{\be}{\begin{equation}}
\newcommand{\ee}{\end{equation}}
\newcommand{\bdm}{\begin{displaymath}}
\newcommand{\edm}{\end{displaymath}}
\newcommand{\bea}{\begin{eqnarray}}
\newcommand{\eea}{\end{eqnarray}}
\newcommand{\no}{\nonumber \\}

\newcommand{\fs}{\; .}
\newcommand{\co}{\; ,}
\newcommand{\al}{&}
\newcommand{\eff}{{e\hspace{-0.1em}f\hspace{-0.18em}f}}

\newcommand{\QCD}{\mbox{\tiny Q\hspace{-0.05em}CD}}
\newcommand{\indR}{\mbox{\tiny R}}
\newcommand{\indL}{\mbox{\tiny L}}
\newcommand{\indV}{\mbox{\tiny V}}
\newcommand{\indP}{{\scriptscriptstyle P}}
\newcommand{\indA}{{\scriptscriptstyle A}}
\newcommand{\indM}{{\scriptscriptstyle M}}
\newcommand{\lvac}{\langle 0|\,}
\newcommand{\rvac}{\,|0\rangle}

\newcommand{\qbar}{\overline{\rule[0.42em]{0.4em}{0em}}\hspace{-0.45em}q}
\newcommand{\ubar}{\overline{\rule[0.42em]{0.4em}{0em}}\hspace{-0.5em}u}

\newcommand{\ph}{\psi}
\newcommand{\ChPT}{ChPT }

\title{On the $1/N$--expansion in chiral perturbation theory}

\author{H. Leutwyler\address{Institute for
  Theoretical Physics, University of Bern, Sidlerstr. 5, CH-3012 Bern,
  Switzerland}%
  \thanks{Work supported in part by Schweizerischer Nationalfonds}}

\begin{document}

\begin{abstract}
In the first part of the talk, I presented a
review of the results for the
light quark masses obtained on the basis of chiral perturbation theory.
As this
material is described in ref.\ \cite{Erice}, the following notes only concern
the second part, which dealt with the behaviour of the effective
theory in the large--$N_c$ limit.\\
{\it Talk given at QCD 97, Montpellier, July 1997}

\end{abstract}

\maketitle

\section{Ward identity}
The low energy properties of QCD are governed by an
approximate,
spontaneously broken symmetry, which originates in the fact that three of the
quarks happen to be light. If $m_u,m_d,m_s$ are turned off, the symmetry
becomes exact: The QCD hamiltonian is then invariant under
independent
rotations of the right- and lefthanded quark fields. For such a theory to
describe the strong interactions observed in nature, it is crucial that
this symmetry is spontaneously broken, the ground state being invariant
only
under the subgroup generated by the charges of the vector currents. There are
theoretical arguments indicating that chromodynamics indeed leads to the
formation of a quark condensate, which is invariant under the subgroup
generated by the vector charges, but correlates the right- and lefthanded
fields and thus breaks chiral invariance \cite{Vafa Witten}. The available
lattice
results also support the hypothesis. In the following I take this
generally accepted picture for granted.

It is easy to see why a nonzero quark condensate implies that the spectrum
of the
theory contains massless particles. Consider the two--point--function
formed with an axial vector
current and a pseudoscalar density,  \bdm A_\mu^a=\qbar
\gamma_\mu\gamma_5\mbox{$\frac{1}{2}$} \lambda^a q\co\hspace{2em}
P^a=\qbar \,i\gamma_5\lambda^a q\co\edm where
$\lambda^0,\,\ldots\,,\,\lambda^8$ is a complete set of hermitean
$3\times 3$ matrices in flavour space, normalized by
$\mbox{tr}(\lambda^a\lambda^b)=2\,\delta^{ab}$. If the quark masses are
dropped, this function obeys the Ward identity
\bea\label{Ward} \al\al\hspace{-2em}\partial^\mu\lvac T
A_\mu^a(x)\,P^b(0)\rvac=\\\al\al\hspace{4em}
-\mbox{$\frac{i}{2}$} \delta(x)\lvac
\qbar\{\lambda^a,\lambda^b\}q\rvac\no \al\al\hspace{4em}
 +\;\mbox{tr}(\lambda^a)\,\lvac T \omega(x)
P^b(0)\rvac\fs\nonumber\eea  The first term on the r.h.s.\ is proportional to
the quark condensate, which in the chiral limit is flavour independent,
$\lvac \qbar_{f'}\,q_f\rvac=\delta_{f'\!f}\,\lvac\ubar u\rvac$.
The second term arises from the anomaly and involves
the winding number density $\omega=
\mbox{tr}_c(G_{\mu\nu}\tilde{G}^{\mu\nu})/(16\pi^2)$.

On account of Lorentz invariance, the Fourier transform of $\lvac T
A_\mu^a(x) P^b(0)\rvac$ is of the form $q_\mu F^{a b}(q^2)$.
For the octet components, where the anomaly term does not
contribute, the Ward identity states that $q^2 F^{a b}(q^2)$ is a constant.
A nonzero quark condensate thus implies that $F^{a b}(q^2)$ contains a
pole at $q^2=0$. The explicit solution of the Ward identity reads
$(a=1,\,\ldots\,,8)$: \bea\al\al\hspace{-2em}
\int\!\! d^4\!x\, e^{i qx}\,\lvac TA^a_\mu(x)P^b(0)\rvac=\no
\al\al \hspace{8em}2\,\delta^{ab}\,
\frac{q_\mu}{q^2}
\,\lvac \ubar u\rvac\fs\nonumber\eea
A pole term of this form arises if and only if the sum over intermediate states
on the left exclusively receives contributions from massless
one--particle states:
The spectrum of QCD must contain an octet of pseudoscalar particles, which
become massless in the chiral limit.

In the
case of the singlet
current, the anomaly spoils current conservation. The Ward identity
does not require the Fourier transform
to contain a pole, but merely implies that the integral
$\int\!d^4\!x\lvac T\omega(x)P^b(0)\rvac$ is determined by the quark
condensate.

\section{Large $N_c$}
Let us now consider the limit $N_c\rightarrow\infty$, at a fixed value of
the renormalization group invariant scale $\Lambda_{\QCD}$ \cite{Large Nc}. In
this limit, the running coupling constant $g$
disappears: $N_c\, g^2$ tends to a constant. The
leading contributions to the connected correlation functions of the quark
currents stem from those graphs that
contain a single quark loop (contributions from graphs with $\ell$ quark
loops are at most of order $N_c^{2-\ell}$). In particular, the
two--point--function considered above is dominated by
graphs with a single quark loop, so that the l.h.s.\
of the Ward identity (\ref{Ward}) is of order $N_c$. On the right, the
contribution
from the quark condensate is also of this order. The graphs relevant for
the anomalous term $\lvac T \omega P^b\rvac$, however, are at most of
order $(N_c)^0$. For large values of $N_c$,
the term that invalidates the
conservation law for the singlet axial current is thus suppressed by one
power of $1/N_c$, so that the argument
given in the preceding section then also applies to the singlet current:
In the limit $N_c\!\rightarrow\!\infty$, the spectrum of
QCD must contain a ninth Goldstone boson. In other words, if the quark
masses $m_u$, $m_d$, $m_s$ are turned off and if the number of colours is
sent to infinity, the mass of the lightest bound state with the quantum numbers
of $P^0\rvac$, the $\eta'$, tends to zero.

To my knowledge, the large--$N_c$ limit represents
the only coherent theoretical explanation of the Okubo-Zweig-Iizuka rule,
whose approximate validity is documented by many examples. It is clear,
however,
that a world containing nine massless strongly interacting particles resembles
the one we live in only vaguely. In particular, the limit strongly distorts the
mass spectrum of the pseudoscalars. We need to account for the terms
generated by the anomaly, even if they tend to zero when $N_c$ becomes large.
This can be done by replacing the limit through an expansion in powers of
$1/N_c$. The method is extensively discussed in the literature and the
leading terms of the effective lagrangian
are known since a long time \cite{Leff U(3)}. More recently, the
expansion in powers of $1/N_c$, momenta and quark masses
was extended to first non--leading leading order \cite{bound,Herrera}. I
wish to discuss some new results obtained on the basis of this approach.

One particular reason for
studying the large--$N_c$ limit in the framework of chiral perturbation theory
is an ambiguity that affects phenomenological
determinations of the quark mass ratios. As pointed out by Kaplan and Manohar
\cite{Kaplan Manohar}, the standard framework only
exploits the symmetry properties of the quark mass term and these remain the
same if the quark mass matrix is subject to the transformation $m'=m+\alpha
\,(m^\dagger)^{-1}\det m$, where $\alpha$ is an arbitrary parameter.
Indeed, the standard effective lagrangian is invariant under the
operation, provided the effective coupling constants are
transformed accordingly. This implies that, beyond leading order,
the quark mass ratios cannot be determined on purely phenomenological grounds.
As noted already in ref.~\cite{Gerard}, however, the ambiguity
disappears in the large--$N_c$ limit,
because the above transformation of the quark mass matrix
violates the OZI rule: The transformation law for the
effective coupling constants shows that the parameter
$\alpha$ is a quantity of order $1/N_c$. Indeed, in the expansion of the
effective
lagrangian introduced below, the ambiguity only shows up at
next-to-next-to
leading order.

The limit $N_c\rightarrow\infty$ enlarges
the symmetry group of the massless hamiltonian from
$\mbox{G}=\mbox{SU(3)}_{\indR}\times
\mbox{SU(3)}_{\indL}\times \mbox{U(1)}_{\indV}$ to
$\mbox{G}=\mbox{U(3)}_{\indR}\times \mbox{U(3)}_{\indR}$, while the subgroup
that leaves the ground state invariant remains the same,
$\mbox{H}=\mbox{U(3)}_{\indV}$. The Goldstone bosons live on the coset space
G/H. For a finite number of colours, G/H = SU(3), while at $N_c=\infty$, G/H =
U(3).
The occurrence of an extra Goldstone boson implies that the standard
chiral lagrangian does not cover the large--$N_c$ limit: A
coherent effective field theory only results if all of the Goldstone
bosons are treated as dynamical variables. The standard framework,
where the effective field $U(x)$ represents an element of SU(3) must be
replaced by one with $U(x)\in\mbox{U(3)}$ \cite{Leff U(3)}.
The unimodular part of
the field $U(x)$ contains the degrees of freedom of the pseudoscalar
octet, while the phase
\bdm\mbox{det}\,U(x)=e^{i\ph(x)}\edm describes the
$\eta^\prime$.

The effective Lagrangian is formed with the field $U(x)$ and its derivatives,
\bdm{\cal L}_\eff\!=\!{\cal L}_\eff (U,\partial U,\partial^2
U,\ldots)\fs\edm The expression may be expanded in powers of $1/N_c$, powers
of momenta and powers of the quark mass matrix $m$.
It is convenient to order this triple series
by counting the three expansion parameters as small quantities of order
$1/N_c\!=\!O(\delta)$,
$p\!=\!O(\sqrt{\delta})$ and $m\!=\!O(\delta)$, respectively \cite{bound}.
The expansion then
takes the form ${\cal L}_\eff={\cal L}^{(0)} + {\cal
L}^{(1)}+\ldots\,$, where the first term is of order $(\delta)^0$,
the second collects the corrections of $O(\delta)$, etc.
The explicit expression
for the leading term reads
\bea\label{L0} {\cal L}^{(0)}\al=\al \mbox{$\frac{1}{4}$}F^2
\langle\partial_\mu U^\dagger\partial^\mu U\rangle\\
\al\al +\mbox{$\frac{1}{4}$} F^2 \langle \chi^\dagger U+
U^\dagger\chi \rangle -\mbox{$\frac{1}{2}$}\tau\ph^2 \co\nonumber\eea
where $\langle X\rangle$ stands for the trace of the $3\times 3$ matrix $X$
and $\chi\equiv 2\,B\,m$.
The three terms represent quantities of order
$N_c\,p^2$, $N_c\, m$ and $(N_c)^0$, respectively.
The first two are familiar from the standard effective Lagrangian. They
involve the pion decay constant
$F\!=\!O(\sqrt{N_c})$ and the constant $B$, which represents a term of order
$(N_c)^0$, related
to the quark condensate. The coefficient $\tau=O[(N_c)^0]$ of the third term is
the topological susceptibility of the purely gluonic theory.
In euclidean
space, this quantity represents the mean square winding number per unit volume,
\bdm \tau=\int\!\!d^4\!x\lvac T\omega(x)\omega(0)\rvac_g\fs\edm

\section{Range of validity of \ChPT}
With the above effective lagrangian, the mass pattern of the pseudoscalar nonet
is readily worked out:
Set $U=e^{i\phi}$ and consider the terms
quadratic in $\phi$,
\bea {\cal L}^{(0)}\al=\al\mbox{$\frac{1}{4}$}F^2
\langle\partial_\mu \phi\partial^\mu \phi\rangle
-\mbox{$\frac{1}{4}$} F^2 \langle \chi \phi^2
\rangle\no\al\al
-\mbox{$\frac{1}{2}$}\tau\langle\phi\rangle^2 +O(\phi^4) \fs\eea
I have expressed the phase of the determinant in terms of the trace
of the field, $\ph=\langle\phi\rangle$. For the off--diagonal components of
$\phi$, this yields the standard mass formulae of current algebra:
\bea\label{mf} M_{\pi^+}^2\al=\al(m_u+m_d)\,B\co\no
    M_{K^+}^2\al=\al (m_u+m_s)\,B\co\\
    M_{K^0}^2\al =\al(m_d+m_s)\,B\fs\nonumber\eea
The diagonal components undergo
mixing -- the levels $\pi^0$, $\eta$ and $\eta'$ repel.
In particular, the
neutral pion is pushed down
and winds up at a mass that is slightly lower than the one of the charged pion
(the effect is proportional to $(m_d-m_u)^2$ and is tiny -- the observed
mass difference is due almost entirely to the e.m.\ interaction).
The prediction for $M_\eta$ and $M_{\eta'}$ depends on the
relative size of the quark masses and the topological susceptibility, which it
is convenient to parametrize through the ratio
\bdm \kappa\equiv \frac{ F^2 B\, m_s}{9\, \tau}\fs\edm
For $\kappa\ll1$, the singlet component of the field effectively decouples,
so that the mass of the $\eta$ is approximately given by the Gell-Mann-Okubo
formula, $M_\eta^2=\frac{1}{3}(4M_K^2-M_\pi^2)$, while $M_{\eta'}^2\simeq
6\,\tau/F^2$. In the opposite limit, $\kappa\gg 1$, where the susceptibility
term becomes irrelevant, the $\eta$ is degenerate with the $\pi$ and
$M_{\eta'}^2=2M_K^2-M_\pi^2$.

The standard framework with $U(x)\in\mbox{SU(3)}$
results if the above effective lagrangian is expanded in powers of
$\kappa$, or equivalently, in inverse powers of $\tau$.
The $\eta'$ then counts among the massive states, which do not show up as
dynamical degrees of freedom, but only manifest themselves indirectly, in the
effective coupling constants. The singlet
component $\ph$ may then be integrated out, generating
a correction of the form $L_7\,\langle \chi^\dagger U-U^\dagger\chi\rangle^2$,
with \cite{GL SU(3)}
\be\label{L7} L_7=-\frac{F^4}{288\,\tau}\,\left\{1+O\left(\frac{1}{N_c}
\right)\right\}\fs\ee

This result gives rise to a paradox \cite{Peris deRafael}:
If the limit $N_c\rightarrow\infty$ is taken at fixed quark masses,
the ``correction term'' proportional to $L_7$ grows with $N_c^2$ and thus
dominates
over the ``leading part'' of the effective lagrangian, which is of order
$N_c$. The reason is that, in this limit, the mass of the $\eta'$ is
comparable to the masses of the remaining pseudoscalars, so that $\ph$
cannot be integrated out. The parameter $\kappa$ becomes large, while the
standard effective lagrangian only covers the region
$\kappa \ll 1$.
Nevertheless, it is perfectly meaningful to consider
the large--$N_c$ behaviour of the effective coupling constants occurring in
the standard framework.
By definition, these are independent of the quark masses. They
may be worked out by treating $m$ as an infinitesimal quantity,
so that the condition $\kappa\ll 1$ is obeyed. The resulting low energy theorem
(\ref{L7}) is not in conflict with general properties of QCD,
but merely shows that, if the limit
$N_c$ is taken at fixed $m$, we are necessarily leaving the region covered
by standard \ChPT.

We may determine the topological susceptibility with the mass of
the $\eta'$. The mass formula for the $\eta$ then takes the form
\bdm
M_\eta^2=m_0^2-\frac{8\,(M_K^2-M_\pi^2)^2}{9\,(M_{\eta'}^2-m_0^2)}\co\edm
where $m_0^2\equiv\frac{1}{3}(4M_K^2-M_\pi^2)$ is the value of $M_\eta^2$ that
follows from the Gell-Mann-Okubo formula. The remainder originates in
the repulsion between $\eta$ and $\eta'$ and
represents an SU(3)--breaking effect of second order.
Inserting the observed masses, we obtain
$M_\eta=494\,\mbox{MeV}$, to be compared with the experimental value,
$M_\eta=547\,\mbox{MeV}$ and
with the Gell-Mann-Okubo formula, which predicts $M_\eta=566 \,\mbox{MeV}$.
The repulsion between the two
levels thus lowers the value of $M_\eta$ by about 70 MeV.
The shift is proportional to
$\kappa$: The mass of the $\eta$ is approximately given by $M_\eta^2\simeq
m_0^2\,(1-\kappa)$. Although the shift
is about four times too large, it does make sense to treat it as a
correction -- the value $\kappa\simeq
\frac{2}{3}M_K^2/M_{\eta'}^2$ is in the range
where standard \ChPT applies.

\section{Wess-Zumino-Witten
term}
The anomalies of the fermion determinant not only equip the $\eta'$ with a
mass. They also explain the lifetime of the $\pi^0$: At leading order of the
low energy expansion, the transitions
$\pi^0\rightarrow \gamma\gamma$, $\eta\rightarrow\gamma\gamma$ and
$\eta'\rightarrow\gamma\gamma$ are described by
the Wess-Zumino-Witten term, which
accounts for the anomalies in the framework of the effective theory and is
proportional to the number of colours. The
piece relevant for the above transitions is given by
\be \label{WZW}{\cal L}_{\mbox{\tiny WZW}}= -
\frac{N_c\alpha}{4\pi}F_{\mu\nu}\tilde{F}^{\mu\nu}\langle Q_e^2\phi\rangle
\co\ee
where $\alpha$, $F_{\mu\nu}$ and $Q_e$ denote the fine structure constant, the
e.m.\ field strength and the quark charge matrix, respectively. I
normalize the corresponding decay rates with $F_\pi$,
\bdm
\Gamma_{P\rightarrow\gamma\gamma}=\frac{\alpha^2M_\indP^3N_c^2}
{576\,\pi^3F_\pi^2} \,c_\indP^2\fs
\edm
The experimental values given by the Particle Data Group \cite{PDG} correspond
to $c_{\pi^0}=1.001\pm 0.036$,
$c_\eta=0.944\pm 0.040$, $c_{\eta'}=1.242\pm0.027$.

At leading order, the standard framework with 8 Goldstone bosons leads to
$c_{\pi^0}=1$,
$c_\eta=3\,\langle Q_e^2\lambda^8\rangle=1/\sqrt{3}$.
The prediction
for the lifetime of the $\pi^0$ is in perfect agreement with the data, but the
result $c_\eta\simeq 0.58$ is too low. Also, this framework does not
shed any light on the value of $c_{\eta'}$.

In the extended effective theory, the octet interferes with the
singlet. At leading order, the corresponding mixing angle is
determined by
\bdm\mbox{tg}\,\vartheta=-\frac{\sqrt{8}\,(M_K^2-M_\pi^2)}
{3\,(M_{\eta'}^2-m_0^2)}\fs\edm
The result for $c_{\pi^0}$ remains the same, but the one for $c_\eta$
is modified by mixing.
Moreover, we now
also obtain a prediction for the $\eta'$ :
\bea c_\eta\al=\al 1/\sqrt{3}\,
(\cos \vartheta-\sqrt{8}\sin\vartheta)\co\no
c_{\eta'}\al=\al 1/\sqrt{3}\,(\sqrt{8}\cos\vartheta+\sin\vartheta)\fs\nonumber
\eea
Numerically, these relations lead to $\vartheta=-20^\circ$,
$c_\eta=1.09$, $c_{\eta'}=1.34$.
The extended lagrangian thus yields a decent approximation for all
three photonic transitions -- the observed amplitudes differ from the
prediction by less than 15 \%.

\section{Georgi's inequality and higher orders}
Above, I have fixed the quark mass ratios with the mass
formulae (\ref{mf}) that follow from the effective lagrangian (\ref{L0}). We
may drop this input and work out the masses $M_\eta$, $M_{\eta'}$,
leaving the ratio
\bdm S=\frac{m_s}{\hat{m}}=\frac{2\,m_s}{m_u+m_d}\edm open.  As
pointed out by Georgi \cite{Georgi,Peris},
the result for $M_\eta/M_{\eta'}$
is smaller than what is observed, quite irrespective of the value of $S$.
The number which results for
$S=(2M_K^2-M_\pi^2)/M_\pi^2=25.9$ is very close to the upper bound, which is
too low by about 10 \%.

Another evident deficiency of the effective lagrangian (\ref{L0}) is that it
leads
to $F_K=F_\pi$, while the observed values are in the ratio $F_K=1.22\,F_\pi$.
Clearly, the higher order terms cannot be neglected at the 10 \% level. In
fact, it was noted already in ref.\ \cite{GL SU(3)} that these
generate
a shift in the mass of the $\eta$ that counteracts the repulsion from the
$\eta'$.

The explicit expression for the terms of order $\delta$ reads
\cite{bound}
\bea\label{L1}\al\al\hspace{-2em}{\cal L}^{(1)}=
L_5\langle\partial_\mu U^\dagger\partial^\mu U(\chi^\dagger U
+U^\dagger\chi)\rangle\\\al\al + L_8 \langle \chi^\dagger U\chi^\dagger
U+h.c.\rangle  + \mbox{$\frac{1}{12}$}
\Lambda_1 F^2\partial_\mu\ph\partial^\mu \ph\no\al\al
+ \mbox{$\frac{1}{12}$} \Lambda_2 F^2 i \ph\langle \chi^\dagger U
-U^\dagger\chi\rangle
+{\cal L}_{\mbox{\tiny
WZW}} \fs\nonumber\eea
I have discarded contributions of the type $(\partial U)^4$, because they do
not contribute to the quantities under discussion. The terms with
$L_5,\,L_8=O(N_c)$ are familiar
from the standard effective lagrangian. Since $L_4$ and
$L_6$ are of order $(N_c)^0$, they only show up in ${\cal L}^{(2)}$
(for a detailed discussion of the large--$N_c$ counting rules
in the framework of the
effective theory, see ref.\ \cite{GL SU(3)}).
Concerning $L_7$, we need to distinguish the coupling constant
$\hat{L}_7$ occurring
in the U(3)--lagrangian from the one appearing in
the standard
framework.
While $\hat{L}_7$ is of order $(N_c)^0$ and does therefore not
show up in ${\cal L}^{(1)}$, the integration over the singlet component of the
effective field gives rise to the additional contribution discussed above, so
that the effective coupling constant relevant for the  SU(3)--lagrangian is
given by
\bdm L_7=-\frac{F^4(1+\Lambda_2)^2}{288\,\tau}+\hat{L}_7\fs\edm

The coupling constants
$\Lambda_1$ and $\Lambda_2$ are of order $1/N_c$ (I have extracted a factor of
$F^2$, so that these constants
are dimensionless\footnote{In ref.\ \cite{bound}, these couplings are
denoted by $K_1=-\frac{1}{12}F^2\Lambda_2$, $K_2=\frac{1}{12}F^2\Lambda_1$.}).
They describe
differences in the dynamics of the octet and singlet components of
the effective field, which arise from violations of the Okubo-Zweig-Iizuka
rule -- in the large--$N_c$ limit, this rule becomes exact.
The term with
$\Lambda_1$ modifies the kinetic energy of the singlet field; the normalization
factor is chosen such that this modification amounts to
$\partial_\mu\ph\partial^\mu\ph\rightarrow
(1+\Lambda_1)\partial_\mu\ph\partial^\mu\ph$. The constant
$\Lambda_2$ affects the interference term between octet and singlet,
$\ph\langle\chi\phi\rangle\rightarrow(1+\Lambda_2)
\ph\langle\chi\phi\rangle$, and also modifies the mass term of the singlet.

\section{Anomalous dimensions}

As is well-known, the dimension of the singlet axial current is anomalous
\cite{Tarrach}.
This implies that the singlet components of the matrix elements
\bdm\lvac\qbar\gamma_\mu\mbox{$\frac{1}{2}$}\lambda^a q|P\rangle
= i p_\mu F^a_\indP\edm
depend on the running
scale: \bdm\mu\frac{dF^0_\indP}{d\mu}=\gamma_\indA\, F^0_\indP\co\hspace{1em}
\gamma_\indA=-\left(\frac{g}{2\pi}\right)^{\!4}\!+O(g^6)\fs\edm
A change in $\mu$ leads to a multiplicative renormalization:
$F^0_\indP\rightarrow Z_\indA\,F_\indP^0$. The problem does not manifest
itself at leading order, where $F_\eta^0=-\sin\vartheta\,F_\pi$, $F_{\eta'}^0=
\cos\vartheta\,F_\pi$, because the anomalous dimension is of order $1/N_c$.
Indeed, the scale dependence is very weak: $F_\indP^0\propto\exp
(4/\beta_0^2L)$, with $L=\ell n( \mu/\Lambda_{\QCD})$, $\beta_0=11-2/3
N_f$.

For the generating functional of QCD to remain invariant under a change of
the running scale, it does therefore not suffice to only
renormalize the external scalar and
pseudoscalar fields.
In addition, we also
need to renormalize the singlet axial field and the vacuum angle.
The symmetry properties of the generating functional ensure that the
renormalization factors of $a_0^\mu(x)$ and $\theta(x)$ are the same.
To simplify
the renormalization group behaviour of the scalar and pseudoscalar external
fields, it is convenient to replace $s(x)$ and $p(x)$ by the combination $m=
e^{i\theta/3}(s+ ip)$, which is invariant
under the transformations generated by the singlet axial charge.
The
generating functional of QCD does remain invariant under a change of the
running scale if the external fields are subject to the transformation
\bea m(x)&\rightarrow & Z_\indM^{-1}\, m(x)\co \no
a^\mu_0(x)&\rightarrow & Z_\indA^{-1}\, a^\mu_0(x)\co \no
\theta(x)&\rightarrow &
Z_\indA^{-1}\, \theta(x)\co\nonumber \eea
where the term $Z_\indM^{-1}$ represents the familiar factor that describes
the scale dependence of the quark masses.

This property of the generating functional is readily translated into
the language of the effective theory.
To maintain the
transformation law $U\rightarrow V_{\indR} U V_{\indL}^\dagger$, we need to
subject
the singlet field $\psi$ to the same multiplicative renormalization as the
vacuum angle, $\psi\rightarrow Z_\indA^{-1}\psi$, while the octet components of
$\phi(x)$ are scale independent. A change in scale thus modifies the
effective coupling constants according to
$B\rightarrow Z_\indM B$, $\tau\rightarrow Z_\indA^2\,\tau$,
$1+\Lambda_1\rightarrow Z_\indA^2\,(1+\Lambda_1)$, $1+\Lambda_2\rightarrow
Z_\indA\,(1+\Lambda_2)$, while $F$, $\chi$, $L_5$ and $L_8$ are scale
independent.

The Wess-Zumino-Witten term (\ref{WZW}) is
invariant under a change of
scale only up to contributions of order $1/N_c$. To arrive at a scale
invariant effective lagrangian, we need to add the term
\be {\cal L}_1^{(2)}=-
\frac{N_c\alpha\Lambda_3}{18\pi}F_{\mu\nu}\tilde{F}^{\mu\nu}
\psi\co\ee
which belongs to ${\cal L}^{(2)}$.
The coupling constant $\Lambda_3$ describes the OZI-violations in the
transitions $P\rightarrow\gamma\gamma$ and is
of order $1/N_c$. It transforms with
$1+\Lambda_3\rightarrow Z_\indA(1+\Lambda_3)$.

At order $\delta^2$, the
effective lagrangian contains further contributions. In particular, it
contains
terms that
contribute to the symmetry breaking in the photonic transition matrix
elements \cite{Moussallam}.
In the following, I ignore this complication, assuming that the symmetry
breaking is dominated
by the interference between the octet and the singlet, which the framework
described here does account for to first nonleading order.

The
discussion below is based in the diploma work of R.\ Kaiser
\cite{Kaiser}. It relies on the
assumption that the expression ${\cal L}^{(0)}+{\cal L}^{(1)}+{\cal L}_1^{(2)}$
represents a decent approximation to the full effective lagrangian. As
mentioned earlier, an ambiguity of
the Kaplan-Manohar type does not show up in this framework: As the
corresponding transformation of the quark mass matrix violates the OZI
rule, it represents a change of the type
$m\rightarrow m\{1+O(\delta^2)\}$, which is beyond the accuracy of our
analysis.

\section{Results}
Since loops
start contributing only at order $\delta^2$, it suffices to evaluate
the tree graphs. For the masses and decay constants, we may again restrict
ourselves to the terms quadratic in $\phi=\sum_a\lambda^a\phi_a$.
While the masses are the square roots of the eigenvalues of this quadratic
form, the decay constants are the entries of the matrix that diagonalizes it:
The eigenstates are given by
$\varphi_\indP
=\sum_aF^a_\indP\,\phi_a$.
For the photonic
transition matrix elements, the above lagrangian implies
\bea\label{C1} \sqrt{3}\,(F_\eta^8\, c_\eta + F_{\eta'}^8\,
c_{\eta'})\al=\al F_\pi\co\\ \label{C2}
\sqrt{3}\,(  F_\eta^0\, c_\eta + F_{\eta'}^0\,
c_{\eta'})\al=\al \sqrt{8}\,F_\pi(1+\Lambda_3)\fs\hspace{1em}\fs\eea
Both of these relations are manifestly scale invariant.

Next, I exploit the fact that the coupling constants $\Lambda_1$ and
$\Lambda_2$ do not affect the part of the lagrangian that is quadratic
in the component $\phi_8$ of the effective field. If the quark mass ratio
$S=m_s/\hat{m}$ is taken as known, we may determine the constants $L_5$ and
$L_8$ with $F_\pi$, $F_K$, $M_\pi$ and $M_K$, so that we arrive at a parameter
free representation of this part of the lagrangian. Comparing the result
with the expression that follows from the representation
$\phi_\indP=F^8_\indP\,\phi_8+\ldots$ for the eigenstates, we obtain
two analogues of the Gell-Mann-Okubo formula:
\bea\label{FM1} \al\al 3\left\{(F_\eta^8)^2+(F_{\eta'}^8)^2\right\}=
4F_K^2-F_\pi^2\co\\ \label{FM2}\al\al
3\left\{(F_\eta^8)^2M_\eta^2+(F_{\eta'}^8)^2M_ {\eta'}^2\right\}=\\
\al\al\hspace{2em} 4F_K^2M_K^2\,\frac{2S}{(S+1)}
-F_\pi^2M_\pi^2\,(2 S-1)\fs\nonumber\eea
Both of these relations are valid to first nonleading order. The first
one implies
\bdm F^8_\eta=\cos\vartheta_8\, F_8\co\hspace{1em}F^8_{\eta'}=\sin \vartheta_8
\,F_8\co\edm
with $F_8=1.28\,F_\pi$. The mixing angle may be determined with eq.\
(\ref{C1}): The observed values of the transition matrix elements
$c_\eta$, $c_{\eta'}$ require $\vartheta_8=-20.5^\circ$. Inserting this in
(\ref{FM2}), we may then determine the quark mass ratio. The result,
$S=26.6$, is remarkably close to the current algebra prediction,
$S=(2M_K^2-M_\pi^2)/M_\pi^2=25.9$.

The coupling constant $L_5$ generates an off-diagonal
part in the kinetic term. Comparing the coefficient of $\partial_\mu \phi_8\,
\partial^\mu\phi_0$ with the one that follows from the representation in
terms of the eigenstates, we obtain
\bea\label{80}\al\al 3\left\{F_\eta^8 F_\eta^0+F_{\eta'}^8
F_{\eta'}^0\right\}
=\\ \al\al \hspace{4em}
-2\sqrt{2}\,(F_K^2- F_\pi^2)\left(1+O(\delta)\right)\fs\nonumber\eea
The relation shows that, at this order of the low energy expansion, the
vectors $(F^8_\eta,F^8_{\eta'})$ and $(F^0_{\eta},F^0_{\eta'})$ are not
orthogonal to one another: We
need to distinguish the mixing angle of the singlet components,
\bdm F^0_\eta= -\sin\vartheta_0\, F_0\co\hspace{1em}F^0_{\eta'}=\cos
\vartheta_0\, F_0\edm
from the one introduced above. The relation (\ref{80}) determines the
difference:
\bdm \sin
(\vartheta_0-\vartheta_8)=\frac{2\sqrt{2}\,(F_K^2-F_\pi^2)}{3\,F_8^2}
\left(1+O(\delta)\right)\fs\edm
I have written the relation in scale invariant form, making use of the
fact that the difference between $F_0$
and $F_8$ is of order $\delta$.
Numerically, the formula yields
$\vartheta_0\simeq-4^\circ$. The term responsible for
the difference between $F_K$ and $F_\pi$ thus also generates a substantial
difference in the two mixing angles.
Since $\vartheta_0$ turns out to be remarkably small,
the state $|\eta\rangle$ is nearly orthogonal to $A^0_\mu\rvac$ -- in
this sense, the $\eta$ is nearly pure octet. To my knowledge, this is a new
result (see the review by Bijnens \cite{Bijnens}; for more recent
discussions of related phenomena, I refer to \cite{Grunberg,Takizawa}).

The relation (\ref{C2}) implies
$F_0=1.25\, F_\pi\,(1+\Lambda_3)$. The numerical value of $F_0$ cannot be
determined
phenomenologically, because it depends on the renormalization scale --
all of the coupling constants
$\tau$, $\Lambda_1$, $\Lambda_2$, $\Lambda_3$ are scale dependent. We may
express the result for these in terms $\Lambda_3$. The numerical
values of the scale invariant
combinations are $\tau/(1+\Lambda_3)^2=(195\,\mbox{MeV})^4$,
$\Lambda_1-2\Lambda_3=0.25$, $\Lambda_2-\Lambda_3=0.28$.
For the remaining coupling constants, we obtain
$F=90.6\,\mbox{MeV}$, $L_5=2.2\cdot 10^{-3}$, $L_7=-0.3\cdot
10^{-3}$, $L_8=1.0\cdot 10^{-3}$.

\section{Discussion}
The sensitivity of the result for the quark mass ratio to the input used for
the decay rates is shown in fig.1.
\begin{figure}[t]
\centering
\mbox{\epsfysize=5.5cm \epsfbox{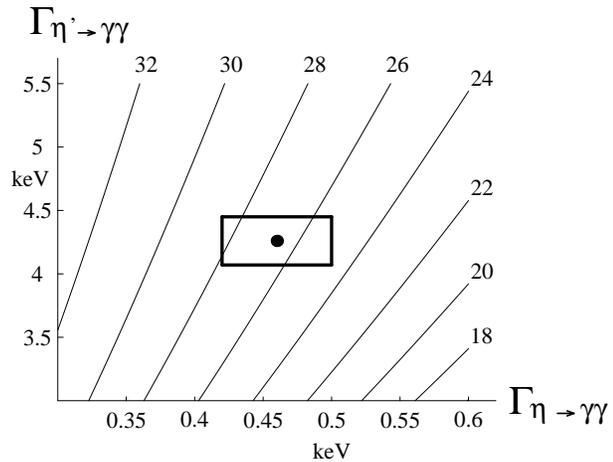} }
\caption{Sensitivity of the quark mass ratio $S=m_s/\hat{m}$ to
the input used for the decay rates $\Gamma_{\eta\rightarrow\gamma\gamma}$ and
$\Gamma_{\eta'\rightarrow\gamma\gamma}$.} \end{figure}
The tilted lines represent constant values of
$S$. The rectangle corresponds to
the experimental errors quoted above.

The calculation described in the preceding section gives rise to another
paradox: According to ref.\
\cite{bound}, the effective lagrangian we are using here leads an upper bound
for $S$. The argument does not invoke the experimental information
about the decay rates and implies that the
current algebra value $S=25.9$ represents an upper limit. Although this is
within the uncertainties of the above result, it is instructive to
identify the origin of the difference.

In ref.\
\cite{bound}, the matrix elements of the $2\times2$ matrices that occur in
the diagonalization procedure are expanded in powers of $1/N_c$ and
only the first two terms are retained. In particular,
$\Delta_N\equiv 24L_5\tau F^{-4}+\frac{1}{2}\Lambda_1-\Lambda_2$ is treated as
a small parameter, because it represents a term of order $1/N_c$. The
contributions from
the OZI-violating coupling constants $\Lambda_1,\Lambda_2$ are indeed small,
but the first term is not: While the susceptibility is related to the mass of
the $\eta'$, $\tau\simeq \frac{1}{6}M_{\eta'}^2 F^2$,
the coupling constant
$L_5$ is dominated by the exchange of scalar resonances,
$L_5\simeq\frac{1}{4}F^2\!/M_{a_0}^2$, so that $\Delta_N\simeq
M_{\eta'}^2/M_{a_0}^2$. Since the two masses are nearly the
same, this estimate yields $\Delta_N\simeq
1$ (indeed, the explicit evaluation of the coupling constants with
the input specified above yields $\Delta_N=1.0$). Despite the fact
that $\Delta_N$ is of order $1/N_c$, it is not
numerically small for $N_c=3$.

The leading contribution to $\Delta_N$
represents a ratio of two terms in the effectie lagrangian: Compare
the term $2 L_5\langle\partial_\mu
\phi\,\partial^\mu\phi \,\chi\rangle$ from ${\cal L}^{(1)}$ to the term
$\frac{1}{4}F^2\langle \phi^2\chi\rangle$ from ${\cal L}^{(0)}$. For the
$\eta'$ matrix elements, the square of the momentum is equal to
$M_{\eta'}^2$, so that the ratio of the two terms is given by
$8 L_5  M_{\eta'}^2/F^2\simeq 48 L_5 \tau/F^4\simeq 2\Delta_N$. Hence the
matrix element of one
of the terms in ${\cal L}^{(1)}$ is about twice as large as the
corresponding matrix element of one of those contained in
${\cal L}^{(0)}$. At first sight, this looks like a desaster for the expansion
we are using here, which treats ${\cal L}^{(1)}$ as a perturbation. It is not
the expansion as such which fails, however.
The ratio is large because the term we picked out from
${\cal L}^{(0)}$ does not represent the main contribution, which arises from
$\frac{1}{2}\tau\ph^2$. Compared to this term, the one from ${\cal
L}^{(1)}$ does indeed represent a small correction. In fact, the results
obtained for the masses and decay constants explicitly show that, throughout,
the first order corrections are reasonably small compared to the
leading terms. In particular, the OZI--violating coupling constants
are small. In this sense, the $1/N_c$--expansion is a
coherent scheme, also in the pseudoscalar channel. As witnessed by the mass
ratio $M_{\eta'}^2/M_{a_0}^2$, it is not true, however, that all
dimensionless quantities of physical interest that vanish for
$N_c\rightarrow\infty$
are numerically small for $N_c=3$. The formal expansion used in ref.\
\cite{bound} in effect treats each one of the terms in
${\cal L}^{(1)}$ as small compared to each one of those in ${\cal L}^{(0)}$ --
this is not the case.

\section{Conclusion}
The simplest form of the large--$N_c$ hypothesis is the assumption
that all dimensionless quantities of physical interest that disappear when
$N_c\rightarrow\infty$ are numerically small for $N_c=3$. In this form, the
hypothesis evidently fails: In the world we live in, the mass ratio
$M_{\eta'}^2/M_{a_0}^2$ is about equal to one, despite the fact
that it represents a quantity of order $1/N_c$. If we wish, we may blame this
on the fact that the topological susceptibility of Gluodynamics happens
to be rather large.

The analysis described here
relies on a weaker form of the large--$N_c$ hypothesis, na\-me\-ly the
assumption that the terms occurring in the effective lagrangian
approximately obey the Okubo-Zweig-Iizuka rule. It implies that those
effective coupling constants
which do not receive contributions from graphs with a single quark loop are
suppressed. At order $p^2$, this hypothesis requires the terms
\bdm\Lambda_1F^2(\partial \ph)^2\hspace{1em}\mbox{and}\hspace{1em}
\Lambda_2F^2i\,\ph\langle\chi U-U^\dagger\chi\rangle\edm to be small
compared to \bdm F^2 \langle\partial U^\dagger\partial U\rangle
\hspace{1em}\mbox{and}\hspace{1em}
F^2\langle \chi U+U^\dagger \chi\rangle\co\hspace{2.1em}
\edm respectively. At order
$p^4$, the same hypothesis
implies that the terms proportional to $L_4$, $L_6$ and $\hat{L}_7$
are small compared to those with $L_5$, $L_8$.
In this form, the $1/N_c$ expansion indeed leads to a coherent
picture for the masses, decay constants and photonic
transitions of the pseudoscalar nonet.

The effective lagrangian used collects all terms of first nonleading order in
the simultaneous expansion in powers of $1/N_c$, momenta and quark masses.
The main limitation of the calculation reported here is that this lagrangian
does not account for the symmetry breaking corrections to the
Wess-Zumino-Wit\-ten term: I am assuming that the symmetry breaking in
the photonic transitions is dominated by the one due to the interference
between the octet and the singlet.

The observed rates for $\eta\rightarrow\gamma\gamma$ and
$\eta'\rightarrow\gamma\gamma$ indicate that the OZI-violations in
these matrix elements are small.
The result obtained for the quark mass ratio $m_s/\hat{m}$ is very close to
the current algebra value. The main effect generated by the
corrections to the well-known leading order lagrangian concerns
$\eta-\eta'$ mixing:
At the order of the low energy expansion considered here, we need to
distinguish two mixing angles. The analysis leads to
a low energy theorem, which states that the difference between the two
is determined by $F_K-F_\pi$. The mixing
angle
seen in the singlet components of the decay constants turns out to be much
smaller than the one in the octet components: $\vartheta_8\simeq - 20^\circ$,
$\vartheta_0\simeq-4^\circ$.

\end{document}